\documentclass[conference]{IEEEtran}
\IEEEoverridecommandlockouts
\usepackage{caption}
\usepackage{graphicx}
\ifCLASSOPTIONcompsoc
  \usepackage[nocompress]{cite}
\else
  \usepackage{cite}
\fi
\usepackage{ragged2e}
\usepackage{cite}
\usepackage{amsmath,amssymb,amsfonts}
\usepackage{mathtools}
\usepackage{algorithmic}
\usepackage{graphicx}
\usepackage{textcomp}
\usepackage{xcolor}
\usepackage{enumitem}
\usepackage{array}
\usepackage{soul}
\usepackage{subcaption}
\usepackage{lipsum}
\usepackage[font=small]{caption}
\newcommand{\PreserveBackslash}[1]{\let\temp=\\#1\let\\=\temp}
\newcolumntype{C}[1]{>{\PreserveBackslash\centering}p{#1}}
\newcolumntype{R}[1]{>{\PreserveBackslash\raggedleft}p{#1}}
\newcolumntype{L}[1]{>{\PreserveBackslash\raggedright}p{#1}}
\DeclarePairedDelimiter{\ceil}{\lceil}{\rceil}

\def\BibTeX{{\rm B\kern-.05em{\sc i\kern-.025em b}\kern-.08em
    T\kern-.1667em\lower.7ex\hbox{E}\kern-.125emX}}
\begin{document}

\title{
FPGA-based AI Smart NICs for Scalable\\ Distributed AI Training Systems
\vspace{-0.3cm}
}

\author{
\IEEEauthorblockN{Rui Ma, Evangelos Georganas, Alexander Heinecke, Andrew Boutros, Eriko Nurvitadhi}
\IEEEauthorblockA{Intel Corporation\\
% E-mail: \{rui.ma, evangelos.georganas, alexander.heinecke, andrew.boutros, eriko.nurvitadhi\}@intel.com}
E-mail contact: eriko.nurvitadhi@intel.com}
\vspace{-0.95cm}
}

\maketitle
\begin{abstract}
Rapid advances in artificial intelligence (AI) technology have led to significant accuracy improvements in a myriad of application domains at the cost of larger and more compute-intensive models.
Training such models on massive amounts of data typically requires scaling to many compute nodes and relies heavily on collective communication algorithms, such as all-reduce, to exchange the weight gradients between different nodes.
The overhead of these collective communication operations in a distributed AI training system can bottleneck its performance, with more pronounced effects as the number of nodes increases.
In this paper, we first characterize the all-reduce operation overhead by profiling distributed AI training.
Then, we propose a new smart network interface card (NIC) for distributed AI training systems using field-programmable gate arrays (FPGAs) to accelerate all-reduce operations and optimize network bandwidth utilization via data compression.
The AI smart NIC frees up the system's compute resources to perform the more compute-intensive tensor operations and increases the overall node-to-node communication efficiency.
We perform real measurements on a prototype distributed AI training system comprised of 6 compute nodes to evaluate the performance gains of our proposed FPGA-based AI smart NIC compared to a baseline system with regular NICs.
We also use these measurements to validate an analytical model that we formulate to predict performance when scaling to larger systems.
Our proposed FPGA-based AI smart NIC enhances overall training performance by 1.6$\times$ at 6 nodes, with an estimated 2.5$\times$ performance improvement at 32 nodes, compared to the baseline system using conventional NICs.
\end{abstract}

%\begin{IEEEkeywords}
%AI training, all-reduce, smart NIC, FPGA
%\end{IEEEkeywords}

\fontsize{9.5pt}{10.9pt}\selectfont

\vspace{-0.2cm}
\section{Introduction}
\label{sec:introduction}

Artificial intelligence (AI) technology is rapidly becoming an essential component of numerous applications.
Recent advances in deep neural networks (DNNs) have led to breakthroughs in a myriad of domains with unprecedented quality of results approaching human-level accuracy.
However, to achieve this level of performance on practical tasks, the compute complexity and memory footprint of DNNs are rapidly increasing at a much faster rate than improvements in our current compute platforms.
For example, state-of-the-art natural language processing models have grown in size from 94M parameters \cite{peters2018elmo} to 540B parameters \cite{chowdhery2022palm} in less than five years.
Therefore, the training of such huge DNNs on massive amounts of data is only feasible on large clusters consisting of many compute nodes (e.g. CPUs, GPUs or custom chips). 
Data parallel training is the most commonly used scheme for distributed learning, in which each node in the system performs forward and backward propagation to calculate weight gradients on a different mini-batch of the training data.
After that, the calculated weight gradients are first aggregated and then sent to all nodes to adjust the model weights based on the weight update rule of the used optimization method (e.g. stochastic gradient descent, Adam~\cite{kingma2014adam}).
This is repeated many times until model accuracy converges and the training process is concluded. 

Efficient scaling of AI training to large distributed systems is challenging.
Ideally, we aim to maximize the utilization of the system's compute nodes performing the key compute-intensive tensor operations in the forward and backward propagation steps of the training process.
However, collective communication operations, such as all-reduce, are required to aggregate the calculated weight gradients from different nodes and update the model weights before starting the next training iteration.
These all-reduce operations consume a portion of the compute node resources, reducing scalability and degrading the overall system performance as less compute resources are now available for executing the core tensor operations.
The problem is further exacerbated for compute nodes integrating increasing numbers of specialized tensor units to deliver higher AI compute throughput, which can be significantly throttled by the all-reduce communication management overheads.

This work proposes offloading the all-reduce operations from the compute nodes of a distributed AI training system to an AI-targeted smart network interface card (NIC) implemented on a field-programmable gate array (FPGA).
This allows the compute nodes to focus primarily on the core compute-intensive tensor operations they are optimized for, while inter-node collective communication is performed simultaneously by the FPGAs.
Moreover, the reconfigurability of FPGAs enables the exploration of different optimizations aiming to improve the overall system communication efficiency.
We enhance our AI smart NIC with a compression engine to perform line rate conversion of single-precision floating-point (FP32) weight gradients into custom block floating-point (BFP) formats similar to that used by Microsoft in their cloud AI accelerators \cite{darvish2020pushing}.
The reconfigurability of the the FPGA fabric enables tuning the BFP format (i.e. block size, mantissa/shared exponent bitwidth) for different workloads to increase the effective NIC bandwidth with minimal effect on model accuracy.
Moreover, our additional AI smart NIC functionality consumes only a small portion of the FPGA resources, making it suitable for adoption in existing FPGA smart NICs \cite{firestone2018azure} or infrastructure processing units (IPUs) that are already targeted for cloud deployments \cite{burres2021intel}.
%% Contributions 
Our contributions in this paper are: 
\noindent
\begin{itemize}[noitemsep,topsep=0pt,leftmargin=2\labelsep]
\item profiling distributed AI training systems to quantify all-reduce communication overheads,
\item implementing an FPGA-based AI smart NIC for accelerating all-reduce and weight gradient compression, 
\item demonstrating that our AI smart NIC enhances training performance by up to 1.6$\times$ in a 6-node prototype system,
\item formulating a detailed analytical performance model showing performance gains up to 2.5$\times$ for larger 32-node systems.
\end{itemize}

\section{Background and Related Work}
\label{sec:background}

\subsection{Distributed AI Training}
Supervised training of DNNs is performed using an application-specific set of labeled training samples $(x_i,y_i)$ where $x_i$ is a data point and $y_i$ is its ground truth label.
The training dataset is split into smaller groups of size $B$ called \textit{mini-batches}.
First, a \textit{forward pass} is performed by passing a mini-batch of inputs through the DNN to generate a prediction $\hat{y}_i$ for each input based on current model weight values.
Then, mean square prediction error $\frac{1}{B}\sum_{i=1}^{B}(y_i - \hat{y}_i)^2$ is used to calculate a \textit{loss function} at the output of the model.
In the \textit{backward pass}, the loss is propagated backwards through all layers to calculate weight gradients (i.e. change in loss with respect to change in weight values).
Finally, weight values are updated based on the calculated weight gradients such that the loss function is minimized.
This process is repeated for all mini-batches in the training set, known as a training \textit{epoch}, and then multiple epochs are performed until accuracy converges.

As DNNs are becoming larger and more compute-intensive, the training procedure becomes practically feasible only when scaled out on a distributed system with many \textit{workers}.
These workers can be any form of compute devices such as CPUs, GPUs, custom accelerators, or even a mix of them.
The most straightforward and commonly used approach for distributed AI training is \textit{data parallelism}.
In this approach, different workers train the whole model using different mini-batches of training data.
Then, they exchange their learning (i.e. weight gradients) after every $N$ training steps to synchronize weight values among themselves using an \textit{all-reduce} operation.
If the trained model is too large to fit in a single worker's memory, another approach known as \textit{model parallelism} is commonly deployed \cite{dean2012large}.
This approach splits the model itself among workers such that each worker is responsible for training a layer (or sub-layer) of the model using all mini-batches.
In this case, intermediate results of the forward and backward passes are transferred from one worker to another.
More recently studied approaches combine data parallelism and model parallelism in a pipelined fashion \cite{harlap2018pipedream}.
This work focuses on data parallel training since it is the most popular and widely applicable approach \cite{li2020pytorch}.

\subsection{All-Reduce Algorithms}
\label{sec:all-reduce}
\begin{figure}[t!]
    \centering
    \includegraphics[width=\linewidth]{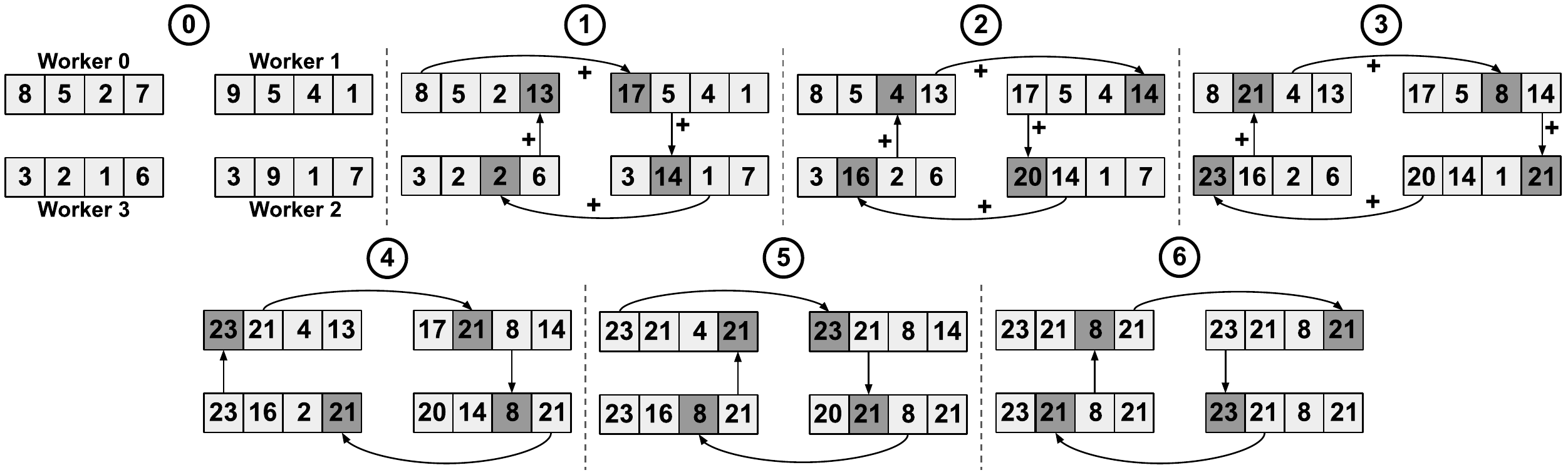}
    \vspace{-0.1cm}
    \caption{Pipelined ring all-reduce with four workers ($w=4$).}
    \label{fig:ring-allreduce}
    \vspace{-0.5cm}
\end{figure}

All-reduce operation used in data parallel training is a collective communication primitive that aggregates data from different workers using an associative operation (e.g. summation) and then distributes the result back to all workers.
It can be implemented in many different approaches such as tree, round-robin, butterfly, and ring topologies \cite{zhao2013sparse, sergeev2018horovod}.
In this work, we implement a pipelined ring all-reduce algorithm which is contention-free and bandwidth optimal at the cost of linear latency scaling with the number of workers \cite{patarasuk2009bandwidth}.
As illustrated in Fig. \ref{fig:ring-allreduce}, each of the $w$ workers sends/receives data to/from one of its neighbors over $2\times(w-1)$ time steps.
In the first $(w-1)$ steps, each worker aggregates the data received from the previous node with its corresponding locally buffered data.
In the remaining steps, workers only store the received final result replacing its corresponding buffered data. 

\subsection{FPGA In-Network Processing}
FPGA-based smart NICs have been commercially deployed to accelerate network functionality on a large scale in the Microsoft Azure cloud since 2015 \cite{firestone2018azure}.
The reconfigurability of FPGAs combines software-like programmability with hardware-like efficiency unlike custom application-specific integrated circuits (ASICs) that lack the flexibility, and general-purpose processors (CPUs or GPUs) that trade efficiency for generality \cite{boutros2021fpga}.
Intel IPUs are another example for FPGA network acceleration; they combine a Xeon CPU with an FPGA fabric to accelerate cloud infrastructure, freeing up the server CPU resources for key datacenter workloads~\cite{burres2021intel}.
These solutions implement high-performance NICs to accelerate application-agnostic network functionality and, unlike our work, do not target AI training.

FPGAs have also been used to implement in-network processing specifically for distributed AI training.
Prior work proposed the implementation of an FPGA-based switch that accelerates weight gradient aggregation and updates on-the-fly as data is transferred between network-connected GPU workers and a host CPU \cite{itsubo2020accelerating}. 
Similarly, iSwitch is another FPGA-based smart switch targeting the more latency sensitive reinforcement learning training \cite{li2019accelerating}.
Gradient aggregations are offloaded from server nodes to a programmable network switch to significantly reduce the number of required network hops and the overall training time. 
Unlike these works, our study focuses on the use of FPGAs as smart NICs instead of switches in distributed DNN training.
Similar to our work, Tanaka et al. \cite{tanaka2020distributed} propose the use of FPGA-based smart NICs to accelerate all-reduce operations in data parallel distributed DNN training.
However, unlike our work, they do not explore the use of the reconfigurable FPGA fabric of their smart NIC for weight gradient compression to increase node-to-node communication efficiency.
In addition, besides real measurements on our 6-node prototype system, we also formulate an accurate performance model to evaluate the scalability of our solution for larger systems with up to 32 workers.

\section{Profiling Distributed AI Training}
\label{sec:cpu-profile}

To quantify the overhead of all-reduce operations in distributed training, we experiment with training feedforward multi-layer perceptrons (MLPs), which represent a major component of many contemporary AI workloads (e.g. recommender systems~\cite{kalamkar2020optimizing}, graph neural networks~\cite{md2021distgnn}).
We choose MLPs as a representative example however, the same training framework is applicable to essentially all other deep learning models.
Our experimental setup is a cluster of Intel Xeon Platinum 8280 28-core CPUs running at AVX512 2.4GHz turbo and 1.8GHz base frequencies.
All CPUs are connected through a network switch via 100 Gbps Ethernet links.
We first deploy a naive solution in which all 28 threads per CPU are performing backward pass tensor operations and when required, one thread triggers an asynchronous all-reduce operation and all the threads (if needed) wait for it to be finished.
This approach exposes the all-reduce latency on the critical path of the training procedure.
Then, we implement a more optimized solution where a number of threads on each CPU are dedicated to handle the all-reduce operation and weight updates in parallel to the remaining threads performing backward pass tensor operations.
This overlap hides most of all-reduce latency behind tensor operations of the backward pass and significantly improves the overall system performance.
Although our setup is based on CPU workers as a proof-of-concept, the same approach and trends apply for network-attached GPU workers (e.g. Nvidia GPUs using GPUDirect RDMA \cite{gpudirect} in DGX systems).

\begin{figure}
    \centering
    
    \begin{subfigure}[b]{\linewidth}
        \centering
        \includegraphics[width=0.9\textwidth]{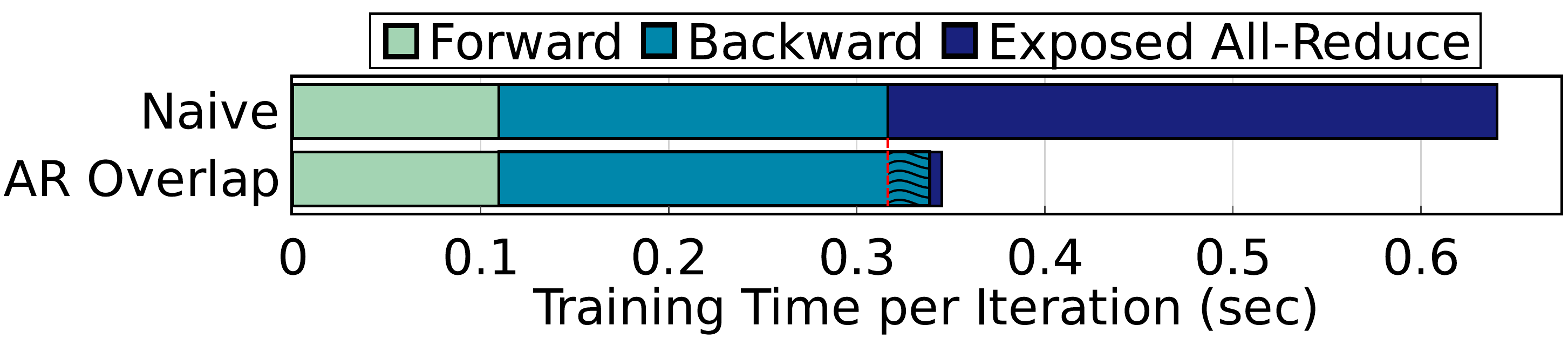}
        \vspace{-0.2cm}
        \caption{}
    \end{subfigure}\\
    \begin{subfigure}[b]{\linewidth}
    \centering
    \includegraphics[width=0.9\textwidth]{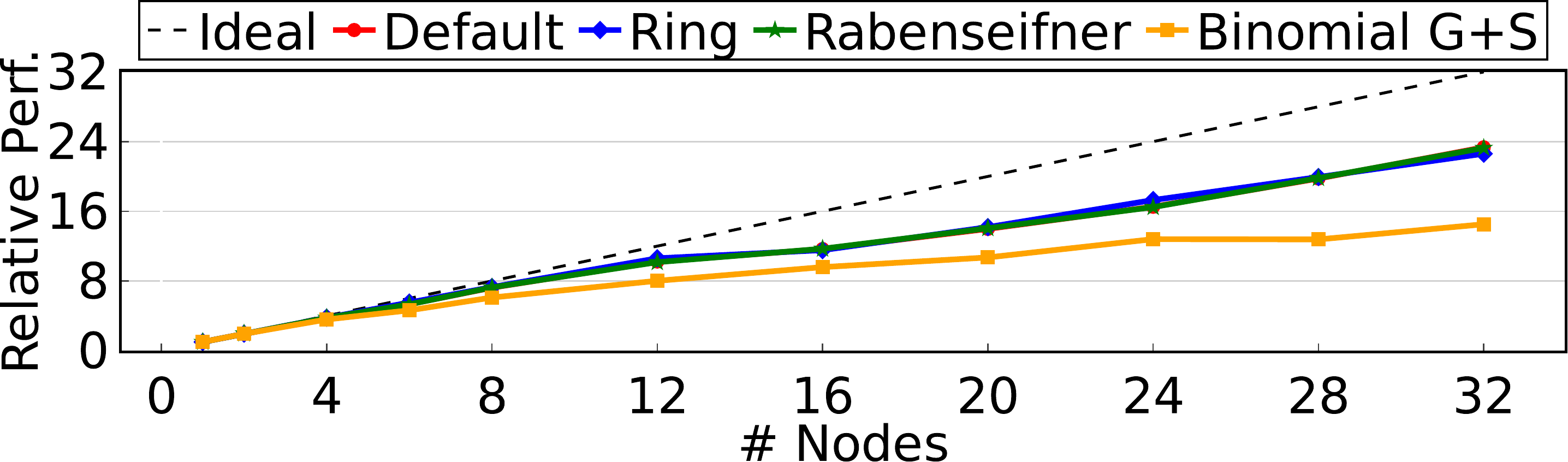}
    \vspace{-0.2cm}
    \caption{}
    \end{subfigure}
    \vspace{-0.7cm}
    \caption{(a) Breakdown of training iteration time for 20-layer MLP on a 6-node system with/without overlapping all-reduce (AR) with backward pass compute. The increase in backward pass time (shaded in black) in the overlapped implementation is due to dedicating a portion of the node resources for managing network communication and executing all-reduce. (b) Scaling of overlapped implementation for different all-reduce schemes.}
    \label{fig:motivation}
    \vspace{-0.5cm}
\end{figure}

Fig. \ref{fig:motivation}a compares the training time per iteration of both approaches for a 20-layer MLP with matrices of size 2048$\times$2048 and mini-batch size of 1792 per node when trained on 6 workers in our CPU cluster.
The exposed all-reduce latency constitutes 51\% of the total execution time of the naive implementation.
For the overlapped implementation, the exposed all-reduce time is 50$\times$ less than the naive approach, resulting in a 1.85$\times$ reduction in the overall training time.
However, the backward pass execution time increases by 11\% (shaded portion in Fig. \ref{fig:motivation}a) since fewer resources are dedicated for the compute-intensive tensor operations.
In this specific experiment, we found that dedicating 2 cores for communication (all-reduce) and 26 cores for compute (backward pass) yields the best performance.
However, this balance between compute/communication cores is workload dependent and needs to be tuned based on number and sizes of layers, mini-batch size and number of workers.
In some cases, the backward pass slowdown can be larger, reducing the overall gains of overlapping communication and compute.

Fig. \ref{fig:motivation}b shows the results of scaling the overlapped implementation to more workers in our CPU cluster compared to the ideal scaling shown by the dashed line.
We experiment with different MPI all-reduce schemes such as the ring, Rabenseifner, and binomial gather/scatter implementations as well as the default setting which employs heuristics to use different algorithms based on message sizes and number of workers \cite{thakur2005optimization}.
The results show that the default, ring and Rabenseifner schemes achieve similar results and are consistently better than the binomial gather/scatter approach.
They also scale well for up to 12 workers, but the gap to ideal scaling gradually increases with the number of workers.
For the rest of our experiments, we use the ring all-reduce scheme which matches what we implement in our system with AI smart NICs.

\section{FPGA-Based AI Smart NIC}

We enhance our distributed training testbed with FPGA-based AI smart NICs to perform in-network all-reduce and hence free up worker resources for the more compute-intensive tensor operations.
In addition, we also leverage the FPGA reconfigurability to implement gradient compression/decompression for more efficient network bandwidth utilization.

\subsection{System Overview and AI Smart NIC Architecture}
\label{sec:system}

\begin{figure}
    \centering
    
    \begin{subfigure}[b]{\linewidth}
        \centering
        \includegraphics[width=0.9\textwidth]{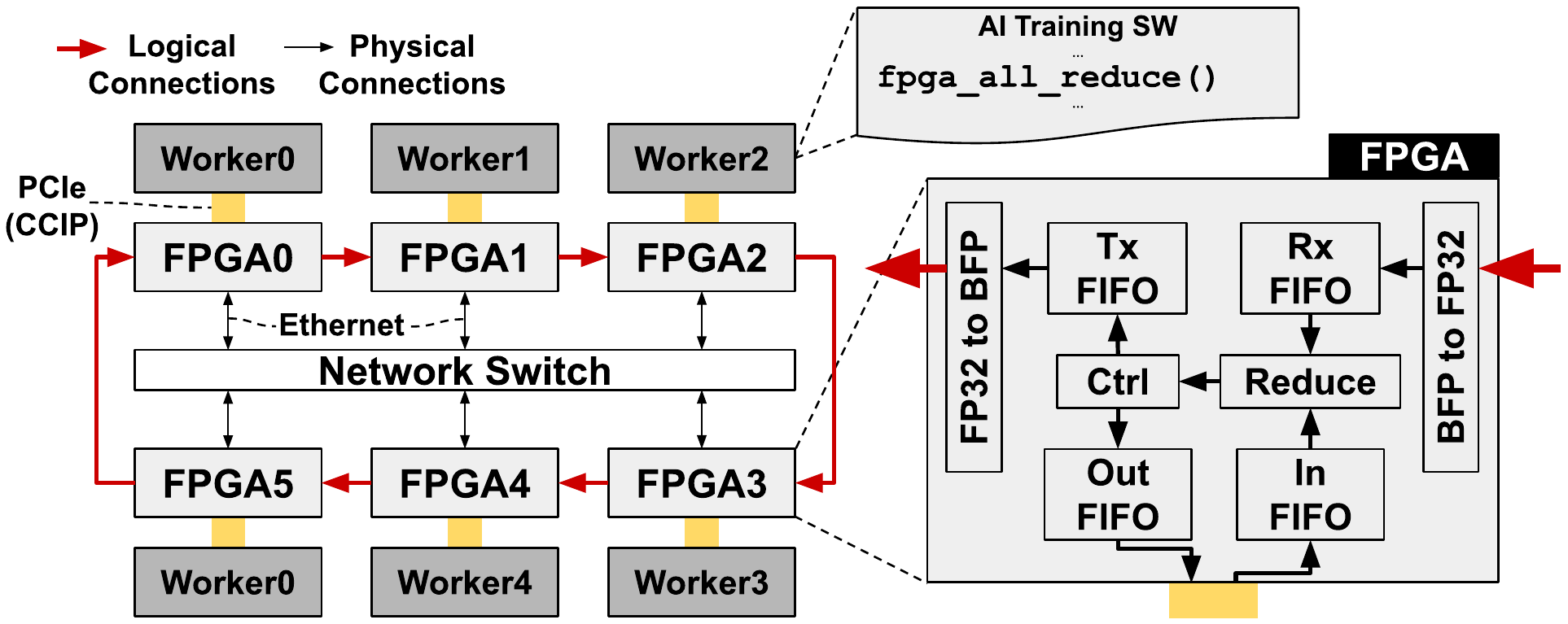}
        \vspace{-0.2cm}
        \caption{}
        \vspace{0.1cm}
    \end{subfigure}\\
    \begin{subfigure}[b]{\linewidth}
        \centering
        \includegraphics[width=\textwidth]{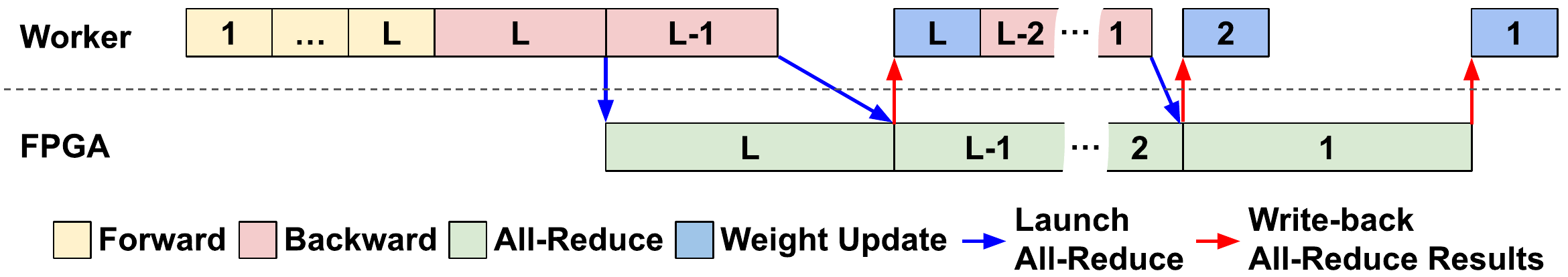}
        \vspace{-0.5cm}
        \caption{}
    \end{subfigure}
    \vspace{-0.5cm}
    \caption{(a) System overview and AI smart NIC architecture, and (b) Example execution trace for $L$-layer MLP training.}
    \label{fig:system}
    \vspace{-0.5cm}
\end{figure}

Fig. \ref{fig:system}a shows an overview of our system enhanced with AI smart NICs.
An FPGA is attached to each worker via PCIe, and all the FPGAs are connected via a network switch.
A ring topology is implemented between the FPGAs on top of the Ethernet layer as shown by the red logical connections in the figure.
Fig. \ref{fig:system}b depicts an example execution trace for training an $L$-layer MLP on our enhanced system.
After each worker executes the forward pass and the backward propagation of the last layer, a thread launches a non-blocking all-reduce request to the FPGA specifying the start memory address and number of the weight gradients to be reduced.
Meanwhile, the workers progress to execute the backward propagation of the next layer, then the same thread launching the all-reduce request blocks execution until the FPGAs finish the all-reduce.
After that, the workers perform weight update and backward propagation of the next layer while the FPGAs are busy with the next all-reduce, which repeats until the complete backward pass is finished.

A simplified diagram of our AI smart NIC architecture is also shown in Fig. \ref{fig:system}a. 
We adopt the same pipelined ring all-reduce algorithm described previously in Section \ref{sec:all-reduce}.
The FPGA first reads weight gradients from its local worker and buffers them in the input FIFO, while the second set of operands of the reduction operation arrives from the previous node over Ethernet and is buffered in the receive (Rx) FIFO.
Once both FIFOs are ready, their contents are dequeued and then reduced using a set of FP32 adders.
Finally, depending on the position of the FPGA and the current step count in the all-reduce ring, a control FSM (Ctrl) decides to steer results to be sent to the next node in the ring via the transmit (Tx) FIFO and/or write it back to the local worker memory as the final all-reduce result via the output FIFO.

\subsection{Block Floating Point Compression}
\label{sec:bfp}

BFP format has shown promising results in reducing the computational complexity and memory footprint of DNNs, with minimal impact on accuracy \cite{drumond2018training, darvish2020pushing}.
In this format, floating point weights or activations are split into small blocks, each of which share the same exponent value among different mantissas. 
The block size and bitwidth of the mantissas and shared exponent are all parameters of the BFP format and represent a tradeoff between efficiency and model accuracy.
Besides the acceleration of all-reduce operations, our AI smart NIC can exploit the reconfigurability of FPGAs to implement other application-specific optimizations such as BFP compression.
As shown in Fig. \ref{fig:system}a, weight gradients are transferred over the network between FPGAs in BFP format;
when recieved by the AI smart NIC, they are decompressed from BFP into FP32 to perform reduction and/or be written back to the worker's memory.
Then, they are compressed again to BFP on the way out when sent over the network to the next node.
In our experiments, we use the BFP16 format from \cite{darvish2020pushing} with 8-bit shared exponents, 7-bit mantissas and a block size of 16 elements, which achieves 3.8$\times$ compression ratio.
However, these parameters can be flexibly adjusted to better suit different workloads making use of the FPGA's flexibility. 

%In addition to offloading the all-reduce, given that our AI NIC is based on an FPGA that is customizable, we propose an optimization scheme where we the AI NIC compress/decompress data sent in the network, hence increasing the effective network bandwidth. We target compression scheme where FP32 data used by CPU is converted to more compact block floating point format in-line by the FPGA, before they are being sent out to other nodes. Training using Block FP has been shown to produce good accuracy [x], hence the Block FP training itself is not new. Here, we showcase novel application of Block FP as a customizable applicable to the proposed AI NIC to optimize its efficiency. There are many other possible customizations that can be applied to the FPGA AI NIC (including other compressoin schemes), which we will study in future work. The block FP scheme studies specifically uses <XX bit mantisa, xx bit exponent, xx block size>. We modified Tensorflow to implement this block FP scheme, and ran experiment to verify  that this precision maintain accuracy for XX network.   

\subsection{Performance Model}
\label{sec:perf-model}

We develop an analytical model to estimate the performance of our system enhanced with AI smart NICs for MLP training.
Assuming that the MLP has $L$ layers each of which has a symmetric weight matrix of dimensions $M_l \times M_l$ for $1 \leq l \leq L$, training mini-batch size is $B$, and the worker's compute throughput is $P_{Worker}$ in FLOPS.
Then the forward ($T_F$) and backward ($T_B$) propagation runtimes for layer $l$ can be calculated as:
\begin{equation*}
    T_{F_l} = \frac{2 \times M_l^2 \times B}{P_{Worker}} ~~~~~~~ T_{B_l} = \frac{4 \times M_l^2 \times B}{P_{Worker}}
\end{equation*}

To implement pipelined ring all-reduce as explained in Sec. \ref{sec:all-reduce} for a system with $N$ nodes, the weight gradients need to be padded (if needed) and evenly split into $N$ chunks.
Therefore, if $b$ is the gradient addition bitwidth, then the total number of bits processed per node in the all-reduce operation of layer $l$ can be calculated as:
\begin{equation*}
\vspace{-0.2cm}
   R_l = b \times N \times \ceil[\bigg]{\frac{M_l^2}{N}}
\end{equation*}

We assume that the AI smart NIC can utilize a portion $\alpha$ of the Ethernet bandwidth $BW_{eth}$\footnote{In all our experiments on a prototype system with 40 Gbps, the value of $\alpha$ was very close to 1.} and we can achieve a compression ratio $\beta$ using the BFP format as discussed in Sec. \ref{sec:bfp}.
Then, the all-reduce runtime is bottlenecked by one of 3 factors: (1) time for transferring data through the ring $T_{ring}$, (2) time for addition $T_{add}$, or (3) time for receiving and writing back data to the worker memory $T_{mem}$.
If we assume PCIe bandwidth is $BW_{pcie}$ and the FPGA addition throughput is $P_{FPGA}$ in FLOPS, we can calculate these 3 factors and the total all-reduce time for layer $l$ as follows.
\begin{equation*}
T_{ring_l} = \frac{R_l \times 2 (N-1)}{N \times \alpha BW_{eth} \times \beta} ~~~~~~~~~~~ T_{add_l} = \frac{R_l \times 2 (N-1)}{N \times P_{FPGA} \times b}
\vspace{-0.2cm}
\end{equation*}
\begin{equation*}
T_{mem_l} = \frac{2 \times Rl}{BW_{pcie}} ~~~~~~~~~~~~~~~ T_{AR_l} = \max(T_{ring_l}, T_{add_l}, T_{mem_l})
\end{equation*}

Finally, we measure the weight update time executed by the worker ($T_U$), which depends mainly on the layer size and memory bandwidth, and linearly scale it for any given layer size.
To calculate the total training iteration time, we use the components $T_{F_l}$, $T_{B_l}$, $T_{AR_l}$ and $T_{U_l}$ to sum up the components of the training trace (similar to that in Fig. \ref{fig:system}b) as follows.
Our results show that our analytical model can estimate system performance within 3\% of the real measurements on our prototype system. 
\vspace{-0.4cm}
\begin{multline*}
T_{total} = \bigg[\sum_{l=1}^{L} T_{F_l}\bigg] + T_{B_L} + \max(T_{B_{L-1}}, T_{AR_L}) + \\
\bigg[\sum_{l=2}^{L-1} \max(T_{U_{l+1}}+T_{B_{l-1}}, T_{AR_{l}})\bigg] + \max(T_{U_{2}}, T_{AR_{1}}) + T_{U_{1}}
\end{multline*}
\vspace{-0.4cm}

\section{Evaluation}

\subsection{Implementation Details}

We implement a prototype 6-node system to evaluate the performance gains of our proposed AI smart NIC.
Each node consists of an Intel Xeon Platinum 8280 28-core CPU attached to an Intel Arria 10 1150 FPGA, similar to that used in the Microsoft Azure smart NIC \cite{firestone2018azure}, via PCIe Gen3x8. 
The FPGAs are connected to a Dell EMC S6100-ON switch via 40 Gbps Ethernet links, on top of which a ring topology is implemented as shown in Fig. \ref{fig:system}a.
We use Intel's IKL for direct inter-FPGA network communication \cite{balle2020inter} and Intel's OPAE stack for CPU-FPGA communication over PCIe~\cite{opae}.
Although our prototype system uses CPU workers, the AI smart NIC enhancement is applicable to systems with GPU (or even custom accelerator) workers as in other prior work \cite{tanaka2020distributed}.

We implement the FPGA-based AI smart NIC in parameterizable Verilog RTL and we use Quartus Prime Pro 17.1 to synthesize, place and route the AI smart NIC logic along with the OPAE and IKL shim.
Table \ref{tab:resources} lists the FPGA resource utilization breakdown for our AI smart NIC, showing that the additional AI-targeted functionality (all-reduce and BFP compression) are very lightweight;
they constitute only 1.2\%, 6.1\%, and 0.5\% of the FPGA's logic, RAM, and DSP resources and thus can be added to existing FPGA smart NICs deployed in production as in the Microsoft Azure cloud \cite{firestone2018azure} with minimal resource overhead.
Although our prototype system uses 40 Gbps Ethernet, there are other FPGA cards that support up to 100 Gbps and 400 Gbps network interfaces.
Therefore, we customize the additional AI smart NIC functionality to match these higher network speeds with 100 Gbps having wider 512-bit interfaces (16 compression and reduction SIMD lanes instead of 8), and 400 Gbps implemented as 4$\times$100 Gbps interfaces.
Our results show that even at 400 Gbps, our proposed additional AI-specific functionality consumes less than 2\%, 9\%, and 5\% of the FPGA logic, RAM, and DSP resources. 

\begin{table}[t!]
    \centering
    \caption{FPGA resource breakdown (ALMs: logic blocks, M20Ks: 20Kb RAMs, DSPs: digital signal processing blocks).}
    \vspace{-0.2cm}
    \begin{tabular}{L{2.3cm} C{1.98cm} C{1.5cm} C{1.5cm}} 
    \hline\\ [-1.8ex]
    \textbf{} & \textbf{ALMs} & \textbf{M20Ks} & \textbf{DSPs} \\ [0.25ex] 
    \hline\\ [-1.8ex]
    OPAE + IKL Shim & 64,480 (15.1\%) & 368 (13.6\%) & 0 (0\%) \\
    All-Reduce & 2,233 (0.5\%) & 46 (1.7\%) & 8 (0.5\%) \\
    BFP Compression & 2,857 (0.7\%) & 120 (4.4\%) & 0 (0\%) \\
    \hline\\ [-1.8ex]
    \textbf{Total} & \textbf{69,570 (16.3\%)} & \textbf{534 (19.7\%)} & \textbf{8 (0.5\%)} \\ [0.25ex]
    \hline
    \end{tabular}
    \label{tab:resources}
    \vspace{-0.6cm}
\end{table}

\subsection{Performance Results}

\begin{figure}
    \centering
    
    \begin{subfigure}[b]{\linewidth}
        \centering
        \includegraphics[width=\textwidth]{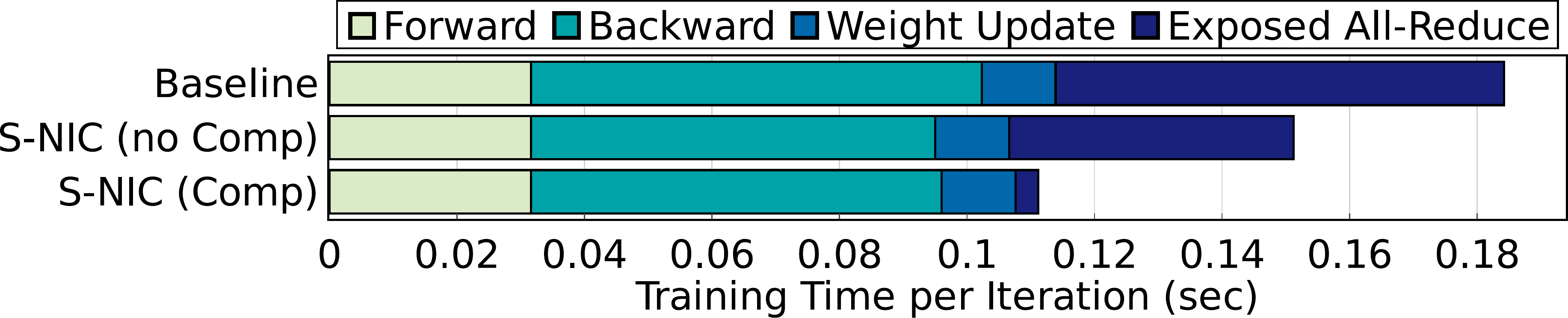}
        \vspace{-0.6cm}
        \caption{}
    \end{subfigure}\\
    \begin{subfigure}[b]{\linewidth}
    \centering
    \includegraphics[width=0.95\textwidth]{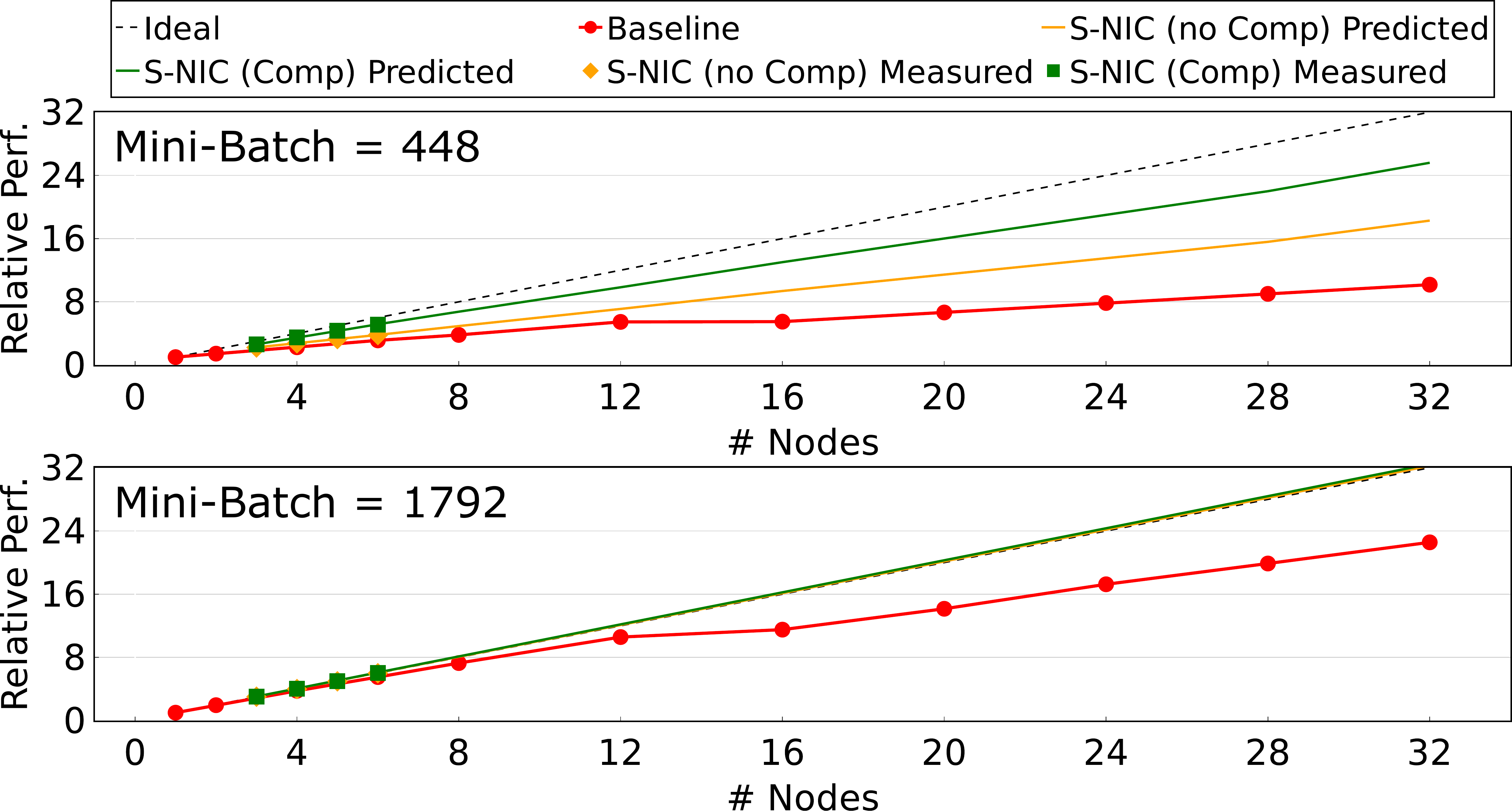}
    \vspace{-0.1cm}
    \caption{}
    \end{subfigure}
    \vspace{-0.7cm}
    \caption{(a) Training time per iteration breakdown of 20-layer MLP and mini-batch of size 448 on a 6-node system for baseline system and smart-NIC-enhanced system with and without BFP compression. (b) Performance scaling for systems with more nodes normalized to a single node compared to ideal scaling (black dashed line).}
    \label{fig:snic-results}
    \vspace{-0.5cm}
\end{figure}

Fig. \ref{fig:snic-results}a compares the training iteration runtime on a 6-node system when using conventional NICs (baseline) and when using our FPGA-based AI smart NIC with and without BFP compression.
For this experiment, we train a 20-layer MLP with 2048$\times$2048 layers and mini-batch size 448.
For the baseline system, we use the optimized implementation overlapping backward propagation compute and all-reduce by dedicating a portion of the workers' resources to network communication management and all-reduce operations, as detailed in Sec. \ref{sec:cpu-profile}.
The results show that the FPGA AI smart NIC, even without using the BFP compression, can reduce the exposed all-reduce time by 37\% and free up the workers' compute resources to focus on the tensor operations of the backward pass reducing it by 10\%.
Overall, this translates to an 18\% reduction in the total training time.
These gains are further increased when deploying the BFP compression on the AI smart NIC;
the exposed all-reduce time and total training time are reduced by 95\% and 40\% respectively compared to the baseline system due to the more efficient utilization of inter-node communication bandwidth.

Fig. \ref{fig:snic-results}b shows how the training throughput (normalized to performance of a system with only 1 worker) scales with increasing the number of nodes for mini-batch sizes of 448 (top) and 1792 (bottom).
The ideal scaling of system performance is showed as a black dashed line for reference.
For the system enhanced with our AI smart NIC, we measure performance for 3, 4, 5 and 6 nodes on our prototype system, then use our analytical model to predict performance for more nodes.
The figures show that measured performance on our prototype system (square and diamond markers) matches very closely the performance predicted by our analytical model within only 3\% error.
For mini-batch size 448, the plot shows that system performance shifts more towards ideal scaling performance (dashed line) when using the FPGA AI smart NIC and when using BFP compression, with performance gains up to 2.5$\times$ and 1.8$\times$ with and without BFP compression respectively.

On the other hand, for mini-batches of size 1792, the system with AI smart NIC achieves ideal scaling as shown in the bottom plot of Fig. \ref{fig:snic-results}b.
In this case, BFP compression provides no additional benefits since for higher batch sizes, the performance is already bottlenecked by the backward propagation compute when using the FPGA AI smart NIC.
Thus, more efficient use of inter-node communication bandwidth via BFP compression does not affect the overall training runtime.
For this batch size, the AI smart NIC increases the overall system performance by 1.1$\times$ and 1.4$\times$ compared to the baseline for systems with 6 and 32 nodes, respectively.
These performance gains are achieved by the AI smart NIC despite using a much lower inter-node network bandwidth of 40 Gbps, compared to the 100 Gbps network used in the baseline system with conventional NICs.

\section{Conclusion}
With the rapid increase in sizes and training data of modern DNNs, distributed training on many-node computing systems becomes inevitable.
Such systems typically rely on all-reduce operations to exchange weight gradients between workers training on different mini-batches.
These operations, despite having low compute intensity, can throttle the overall system performance since a portion of the compute resources of each node is dedicated for performing reduction operations and managing network communication.
To alleviate this bottleneck, we propose an FPGA-based AI smart NIC to accelerate all-reduce operations and free up node compute resources for more compute-intensive backpropagation tensor operations.
The reconfigurable nature of FPGAs also enable custom application-specific optimizations such as BFP compression for more efficient inter-node network communication.
We build a 6-node prototype system demonstrating that our proposed AI smart NIC can reduce the overall training time by 40\%, while adding only 1.2\%, 6.1\%, and 0.5\% additional logic, RAM and DSP resources to existing FPGA smart NICs for the AI-specific functionalities. 
We also formulate a detailed model to estimate performance for systems with more nodes, showing that our AI smart NIC can potentially increase training performance by up to 2.5$\times$ in 32-node systems. 

\vspace{-0.2cm}
\bibliography{references}
\bibliographystyle{ieeetr}

%\end{IEEEbiography}


%\newpage



%\vfill


\end{document}